# Spin-Triplet Vortex State in the Topological Superconductor $Cu_xBi_2Se_3$

Pradip Das, Yusuke Suzuki, Masashi Tachiki and Kazuo Kadowaki

*Institute of Materials Science and Graduate School of Pure & Applied Sciences, University of Tsukuba, 1-1-1, Tennodai, Tsukuba, Ibaraki 305-8573, Japan*

Abstract: We report on the observation of bulk superconductivity from dc magnetization measurements in a cylindrical single crystal of $Cu_xBi_2Se_3$. The magnitude of the magnetization in the Meissner state is very small and the magnetic-field dependence of the magnetization just above the lower critical field $H_{c1}$ is very different from those of usual type-II superconductors. We studied the character of the vortex state theoretically in a spin-triplet pairing superconductor and compared it with the experimental results. The results showed that, the superconductivity observed in $Cu_xBi_2Se_3$ is consistent with the spin-triplet pairing superconductivity with odd parity. We also observed a rapid relaxation phenomenon of the superconducting diamagnetism.

Topological insulators are materials with a bulk-insulating gap that exhibit quantum-Hall-effect-like behavior in the absence of a magnetic field. The experimental as well as theoretical studies [1-9] have shown that $Bi_2Se_3$ has a large band gap (~ 0.3 eV) and a single surface Dirac cone associated with the topologically protected surface state vis-à-vis time reversal symmetry. An early report described the quantum spin Hall effect observed in an Hg-Te system, [10-11] after that topological insulator was reported in Bi-Sb alloy. [1] Since then, topological properties were reported in several doped or undoped $Bi_2Se_3$ and $Bi_2Te_3$ compounds, *e.g.*, $Ca_xBi_{2-x}Se_3$ [12], $Mn_xBi_{2-x}Te_3$ [13], $Cu_xBi_2Se_3$ [14]. In these Bi related topological insulators, the large spin-orbit coupling plays an essential role. Among these materials, $Cu_xBi_2Se_3$ is of particular interest for theoretical as well as experimental scientists because of the signature of superconductivity found at low temperatures by Hor *et. al.* [14-15]. A Pressure-induced superconducting state was also reported in $Bi_2Te_3$ by Zhang and Einaga *et. al.* [16-17]. Their results strongly suggest that the superconducting state at low temperature points toward the co-existence of the Cooper pair with the topological time-reversal protected surface state. This brings us to a question concerning the nature of superconductivity as well as the relation between two distinct properties of material: superconductivity and a topological insulator state. The origin of the observed superconductivity in a topological insulator intimately relates to a fully gapped state in the bulk along with the gapless surface Andreev bound states[15]. Some theories indicate the existence of the Majorana state bound across the grain boundary in the sample due to the co-existence of superconductivity and a topological insulator [4-6]. While such a scenario is predicted in the literature, some ideas suggest the odd-parity superconducting state in this sample [15, 18, 19]. Contrary to the latter, Sato *et. al.* predict an even-parity superconducting state [20]. Although Hor *et. al.* reported [14] the observation of superconductivity from a temperature scan of magnetization and resistivity, such studies in the literature are very limited and as yet there is no detailed picture associated with a field scan.

In this paper, we report on the growth and magnetization properties of single crystalline $Cu_xBi_2Se_3$. Detailed of magnetization measurements were carried out for several single crystals for $H // c$ and $H \perp c$. Copper intercalation in the Van der Waals gap between $Bi_2Se_3$ is supposed to be responsible for the reported observations [14] of the induced superconductivity in $Cu_xBi_2Se_3$. In the present study, we observe the bulk superconductivity in $Cu_xBi_2Se_3$ topological insulator. We believe that the nature of the vortex state in this superconductor is exotic and different from conventional or cuprate high-$T_c$ superconductors. We propose that the observed rapid magnetic field evolution of magnetization signal is characteristic of the odd-parity, i.e., the spin-triplet-superconducting state due to the strong spin-orbit coupling in the superconductor. We also observe anisotropy and rapid relaxation in our sample.

$Bi_2Se_3$ and $Cu_{0.15}Bi_2Se_3$ single crystals were prepared by reacting pressed pellets of a thoroughly mixed powder of Bi, Se, and Cu sealed in an evacuated quartz tube at 1123 K for 24 hours and then cooled at a rate of 2.5 K/min. to 893 K following liquid nitrogen quenching. The silver colour platelet crystals were obtained and easily cleaved off from the ingots of the obtained single crystal. Small pieces of the single crystals were subjected to x-ray diffraction, using a PANalytical x-ray diffractometer with monochromatic Cu Kα radiation, to check the phase purity and to estimate the lattice constant values. No trace of any secondary phase was seen in the x-ray diffreactometer, indicating the sample is phase-pure (see Figure 1 (a)). Good crystallinity is proved by the sharp *c*-axis (00l) Bragg reflections in x-ray diffraction. We noticed a significant increase in the c-axis lattice parameter from c= 28.647 (2) Å for $Bi_2Se_3$, to c= 28.697 (2) Å for $Cu_{0.15}Bi_2Se_3$. . For the magnetization measurement, we have chosen a cylindrical shaped crystal (diameter ~ 6 mm and height ~ 1mm) of $Cu_{0.15}Bi_2Se_3$ single crystal. The dc magnetization was measured with a SQUID magnetometer (Quantum Design Inc., USA).

Figure 1(b) shows representative plots of ZFC (zero field cooled) & FCC (field cooled cooling) temperature variation of dc magnetization for the $Cu_{0.15}Bi_2Se_3$ crystal with *H//* [001] for fixed fields of 10 Oe, 50 Oe, and 700 Oe with temperature. The ZFC data at 10 Oe show a clear diamagnetic signal from the normal to the superconducting transition at ~ 3.6 K. The normal state is diamagnetic with $\chi_{dia} = -2.2 \times 10^{-6}$ emu/mol without significant temperature dependence up to at least room temperature. This behavior is nearly the same for the mother compound $Bi_2Se_3$ as well. The superconducting signal down to 2 K is not fully saturated, indicating that the whole sample may not be superconducting, but a very small fractional part of the obtained sample is superconducting down to the lowest achievable temperature in our experiment. The ZFC and FCC curves do not follow the same path, showing the typical behavior for type-II superconductors with pinning. To verify whether a thermal hysteresis exists across the superconducting region, we measured the magnetization in field-cooled- warm (FCW) mode also. Figure 1(c) shows representative plots of the temperature variation of the FCW magnetization curves along with the corresponding FCC curves for 50 Oe. We found that the irreversibility between $M_{FCC}$ and $M_{FCW}$ is negligible.

Figure 2 (a) displays typical isothermal magnetization hysteresis loops (only part of the first quadrant is shown for clarity) recorded at 2 K with *H//* [001] and $H \perp$ [001]. The initial portion of the *M-H* curve can be utilized to determine the lower critical field by peaking up the turnaround field. At 2 K, $H_{c1}$ in the given cylindrical sample is ~ 5 Oe (see upper inset of Fig. 2 (a)). It can be noted that the *M-H* loop is reversible above 1500 Oe, and at very high field *i.e.* in the normal state the signal is still diamagnetic consistent with isofield measurements. The in-plane measurement at 2 K is similar to the one for the out-of-plane measurement, and there is very little change at low fields around the turnaround field value (~ 5 Oe, see the upper inset of Fig. 2(a)).

As seen in Fig. 2 (a), when the external magnetic field is increased, the magnetization sharply increases in a small range of the magnetic field just above the lower critical magnetic field $H_{c1}$. Then, the magnetization gradually increases with increasing field and finally merges in the normal state diamagnetic curve at $H_{c2}$. This behavior is very different from that of usual type-II superconductors. For the topological superconductor $Cu_xBi_2Se_3$, Liang Fu and Erez Berg [18] considered that the superconductor has a full pairing gap in the bulk and a gapless Andreev bound state at the surface. Through the study of the pairing symmetry of the superconductor, they found that a novel spin-triplet pairing with odd parity is favored by the strong spin-orbit coupling in the superconductor. On the basis of this theory, we investigated the character of the vortex state in the present superconductor.

The vortex current induces a non-uniform magnetic field $h(r)$ in the vortex. This magnetic field polarizes the spins of the spin-triplet pairs and produces a non-uniform spin magnetization $m(r)$ in the crystal. Along with the vortex current, the spin magnetization contributes to the magnetic flux $b(r)$ in the superconductor. The magnetic flux is given by a sum of the magnetic field and the spin magnetizations as

$$b(r)= h(r) +4\pi m(r ). \qquad (1)$$

The total magnetic flux of a single vortex is quantized. The flux quantization leads to the following situation: the vortex current is drastically affected by the spin magnetization and the current inversion occurs in some portion of the vortex. The current inversion causes the attractive force between the vortices. The magnetization curve is strongly influenced by the attractive interaction. This effect brings about a drastic increase of the magnetization just above $H_{c1.}$ The phenomena mentioned above are formulated as follows.

According to Ref. [21] the interaction energy between the vortices whose centers are at the origin and a position $r$ is given by

$$\frac{\phi}{8\pi}h(r), \qquad (2)$$

where $\phi$ is the quantized magnetic flux. To calculate $h(r)$, we set up an equation, following the calculation procedure given in Ref.[21]. The obtained integral-differential equation, which contains valuables $h(r)$, $m(r)$, and the phase of the superconducting order parameter $\varphi$ $(r)$, is similar to Eq. (3.2) in Ref. [21]. The phase for a single vortex fulfills the equation

$$\nabla\times\nabla\,\varphi(r) = 2\pi\mathbf{e}\,\delta(r), \qquad (3)$$

where $\mathbf{e}$ is a unit vector along the magnetic field. The spin magnetization $m(x)$ is connected with the magnetic field $h(y)$ by using the nonlocal susceptibility $\chi(x- y)$ as

$$m(x) = \int dy\,\chi(x-y)h(y). \qquad (4)$$

The nonlocal susceptibility for the present system has not been obtained. Here we use tentatively the susceptibility for the paramagnetic state of the ferromagnetic superconductor used in Ref. [21]. We obtained the values of *h* (*r*) by numerically solving the integral-differential equation together with Eqs. (3) and (4). The result is schematically shown in Fig. 2(b). The dotted green line shows the magnetic field of the spin-singlet superconductor. The magnetic field *h* (*r*) is always positive for the entire region of *r*, indicating that the interaction force between the vortices is always repulsive. In this case, the magnetization curve is of usual type-II superconductors. On the other hand, the dashed red line in Fig.2 (b) shows the magnetic field for the triplet-pairing superconductor. In this case, the negative part of *h* (*r*) appears in its tail and exerts an attractive force between the vortices. The attractive force helps the vortices to enter into the crystal although the vortex density is small at the external magnetic field just above $H_{c1.}$ This fact explains the experimental results that the magnetization drastically increases in the magnetic-field range, as seen in Fig. 2(a). When the external magnetic field is increased furthe, the vortex density increases and the distance between the vortices becomes shorter. In this regime, the repulsive interaction begins to work as seen in Fig. 2(b), and then the increase of the magnetization becomes gradual in agreement with the experimental results shown in Fig.2 (a). The results discussed above strongly support the idea that the superconductivity in the topological superconductor $Cu_xBi_2Se_3$ is due to the spin-triplet pairing with odd parity, not singlet pairing. The detail of the theoretical calculation will be published elsewhere. Here, we wish to point out that the incomplete Meissner state in the sample may be due to the accumulation of flux in the conducting surface state which is promptly expelled from the bulk superconducting sate. Figure 3 shows plots of temperature variation of ZFC dc magnetization with *H*// [001] and $H \perp$ [001] for fixed fields 10 Oe, 50 Oe and 700 Oe. The small separation between the magnetization signal in the ZFC data along with the observed similarities in the *M*-*H* data between the in-plane and out-of plane are the signature of small anisotropy even in the layered structure of the $Cu_xBi_2Se_3$ system, providing the strong evidence that the superconductivity is in bulk nature.

Figure 4(a) and (b) displays the magnetic moment versus time plot at 2 K with *H*// [001] and $H\perp$[001] for different filed values, as indicated in the figure. Magnetic relaxation experiments have been carried out as a function of applied magnetic field by choosing a particular field value where the sample is in the superconducting region in the field-temperature phase space. In our measurements, the sample was cooled from the normal state in zero field condition to the chosen temperature (2 K) and the magnetic field was switched on at a constant rate of 100 Oe/s. The measurement is started when the field reaches the desired value and is continued for 1 hour. We observe a rapid relaxation of the magnetization signal, and after 100 second the magnetization signal is almost stable for the duration of the experiment. The observed fluctuations in the data are the hallmark of the rapid vortex motion in that temperature and field regime. Figures 4 (c) and (d) display the magnetic moment versus logarithmic time plot (reploted from fig. 4 (a) and (b)) at 2 K with *H//* [001] and *H*⊥[001] for different filed values, as indicated in the figure. We observed logarithmic decay of the magnetization signal prominently.

In conclusion, we have presented the experimental data pertaining to the fact that the difference between the magnetization curves parallel and perpendicular to the *c*-axis is very small, and the superconductivity occurs in the bulk of the single crystal of $Cu_xBi_2Se_3$. From the experimental and theoretical studies of the magnetic-field dependence of the magnetization just above $H_{c1}$, we conclude that the vortices are those in the superconducting state of spin-triplet pairing. The rapid relaxation of the diamagnetic magnetization indicates the flexible motion of the vortices.

This work has been supported in part by CREST-JST (Japan Science and Technology Agency), WPI (World Premier International Research Center Initiative)-MANA (Materials Nanoarchitectonics) project (NIMS), and Strategic Initiative category (A) at the University of Tsukuba.

## References

Contact E-mail:
kadowaki@ims.tsukuba.ac.jp
tachiki@ims.tsukuba.ac.jp
das@ims.tsukuba.ac.jp

[1] D. Hsieh, D. Qian, L. Wray, Y. Xia, Y. S. Hor, R. J. Cava, and M. Z. Hasan, Nature (London) **452**, 970 (2008).

[2] J. Moore, Nature Phys. **5**, 378 (2009).

[3] M. Z. Hasan and C. L. Kane , Rev. Mod. Phys. **82**, 3045 (2010).

[4] L. Fu and C. L. Kane, Phys. Rev. Lett. **100**, 096407 (2008).

[5] F. Wilczek, Nature Phys. **5**, 614 (2009).

[6] L. Fu and C. L. Kane, Phys. Rev. Lett. **102**, 216403 (2009).

[7] D. Hsieh, Y. Xia, D. Qian, L. Wray, F. Meier, J. H. Dil, J. Osterwalder, L. Patthey, A.V. Fedorov, H. Lin, A. Bansil, D. Grauer, Y. S. Hor, R. J. Cava, and M. Z. Hasan, Phys. Rev. Lett. **103**, 146401 (2009).

[8] Y. Xia, D. Qian, D. Hsieh, L. Wray, A. Pal, A. Bansil, D.Grauer, Y. S. Hor, R. J. Cava, and M. Z. Hasan, Nature Phys. **5**, 398 (2009).

[9] Y. L. Chen, J. G. Analytis, J.-H. Chu, Z. K. Liu, S.-K. Mo, X. L. Qi, H. J. Zhang, D. H. Lu, X. Dai, Z. Fang, S. C. Zhang, I. R. Fisher, Z. Hussain, Z.-X. Shen, Science **325**, 178 (2009).

[10] L. Fu and C. L. Kane, Phys. Rev. B **76**, 045302 (2007).

[11] S. Murakami, New J. Phys. **9**, 356 (2007).

[12] Y. S. Hor, A. Richardella, P. Roushan, Y. Xia, J. G. Checkelsky, A. Yazdani, M. Z. Hasan, N. P. Ong, and R. J. Cava, Phys. Rev. B **79**, 195208 (2009).

[13] Y. S. Hor, P. Roushan, H. Beidenkopf, J. Seo, D. Qu, J. G. Checkelsky, L. A. Wray, D. Hsieh, Y. Xia, S.-Y. Xu, D. Qian, M. Z. Hasan, N. P. Ong, A. Yazdani, and R. J. Cava, Phys. Rev. B **81**, 195203 (2010).

[14] Y. S. Hor, A. J. Williams, J. G. Checkelsky, P. Roushan, J. Seo, Q. Xu, H.W. Zandbergen, A. Yazdani, N. P. Ong, and R. J. Cava, Phys. Rev. Lett. **104**, 057001 (2010).

[15] L. Andrew Wray, Su-Yang Xu, Yuqi Xia, Yew San Hor, Dong Qian, Alexei V. Fedorov, Hsin Lin, Arun Bansil, Robert J. Cava and M. Zahid Hasan1, Nature Phys. **6**, 855 (2010).

[16] J. L. Zhang, S. J. Zhang, H. M. Weng, W. Zhang, L. X. Yang, Q. Q. Liu, S. M. Feng, X. C. Wang, R. C. Yu, L. Z. Cao, L. Wang, W. G. Yang, H. Z. Liu, W. Y. Zhao, S. C. Zhang, X. Dai, Z. Fang, and C. Q. Jin, Proceedings of the National Academy of Sciences **108**, 24 (2011).

[17] M Einaga1, Y Tanabe1, A Nakayama, A Ohmura, F Ishikawa, and Yuh Yamada, J Phys Conf Ser. **215**, 012036 (20110).

[18] L. Fu and E. Berg, Phys. Rev. Lett. **105**, 097001 (2010).

[19] Masatoshi Sato, Phys. Rev. B **81**, 220504 (2010).

[20] Masatoshi Sato, Yoshiro Takahashi, and Satoshi Fujimoto, Phys. Rev. B **82**, 134521 (2010).

[21] M. Tachiki, H. Matsumoto, and H. Umezawa, Phys. Rev. B **20**, 1915 (1979).

**Figure Captions**

**Fig. 1.** (Color online) Panel (a) shows x-ray-diffraction pattern of $Bi_2Se_3$ and $Cu_xBi_2Se_3$ (x=0, 0.15) at room temperature. Panel (b) shows zero field cooled (ZFC) and field cooled cooling (FCC) mode magnetization data as a function of temperature, recorded at fields of 10 Oe, 50 Oe, and 700 Oe for *H//c*. Panel (c) shows a comparison of the temperature dependence of the magnetization in field-cooled cooling mode, and field-cooled warming (FCW) up mode at 50 Oe field.

**Fig. 2.** (Color online) (a) Main panel shows a portion of the dc magnetization hysteresis curve in a single crystal of $Cu_{0.15}Bi_2Se_3$ for *H//c* and $H\perp c$ at 2 K. Upper inset shows the expanded portion of the M-H curve across the low field regime. Lower inset shows the portion of the dc magnetization hysteresis curve up to 4 Tesla field for $H\perp c$ at 2 K. (b) Schematic plot of magnetic field *h* (*r*), for the singlet (dotted green line) and triplet (dashed red dotted line) pairing superconductor.

**Fig. 3.** (Color online) Zero field cooled (ZFC) mode magnetization data as a function of temperature, recorded at fields of 10 Oe, 50 Oe, and 700 Oe for *H//c* and $H\perp c$.

**Fig. 4.** (Color online) Panels (a) and (b), show the time variation of the in-plane and out-of-plane magnetization at fields, respectively, as indicated. Panels (c) and (d) show the logarithmic time variation of the in-plane and out-of-plane magnetization at fields, respectively, as indicated.

**Figures**

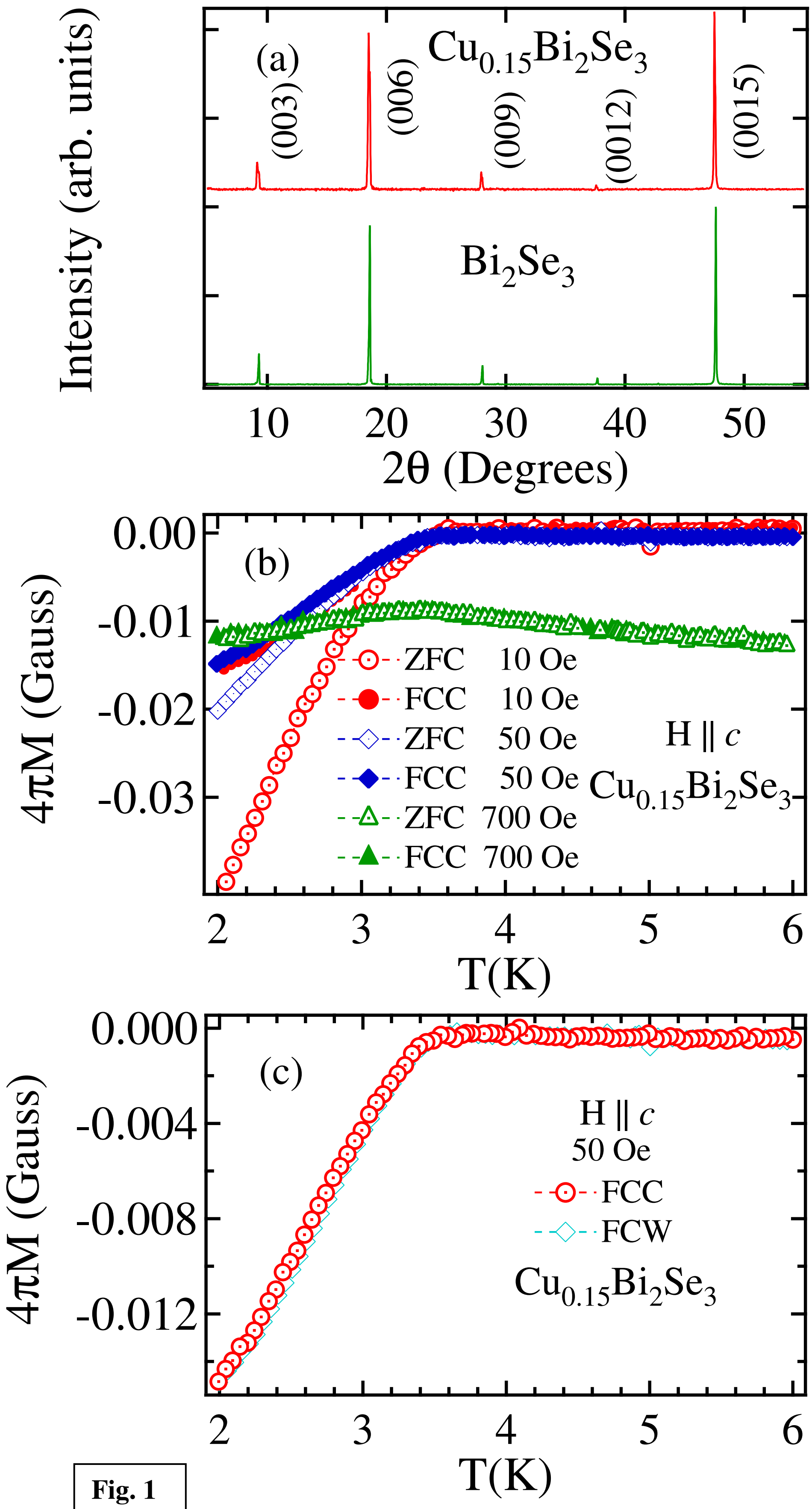

**Fig. 1**

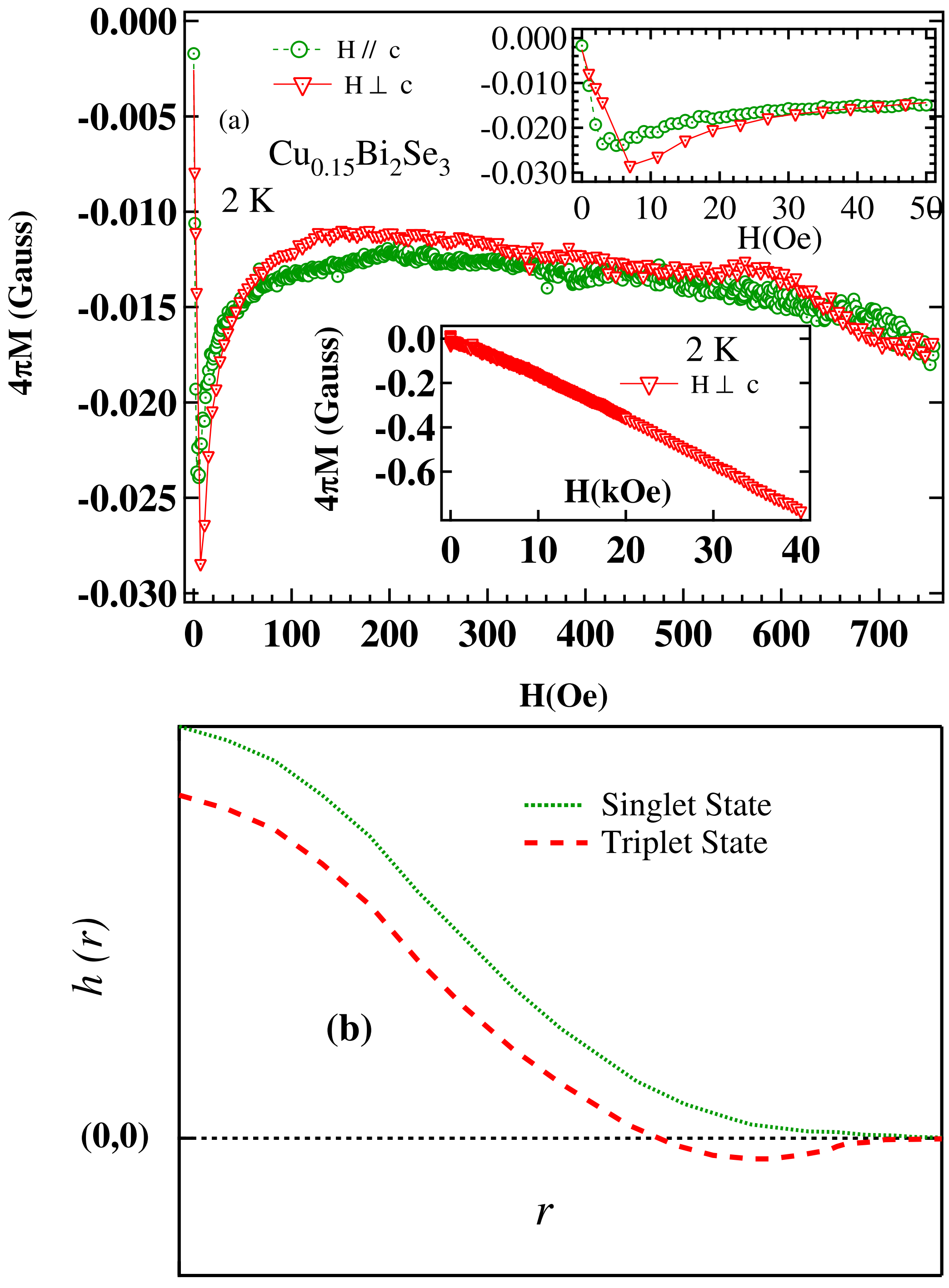

**Fig. 2**

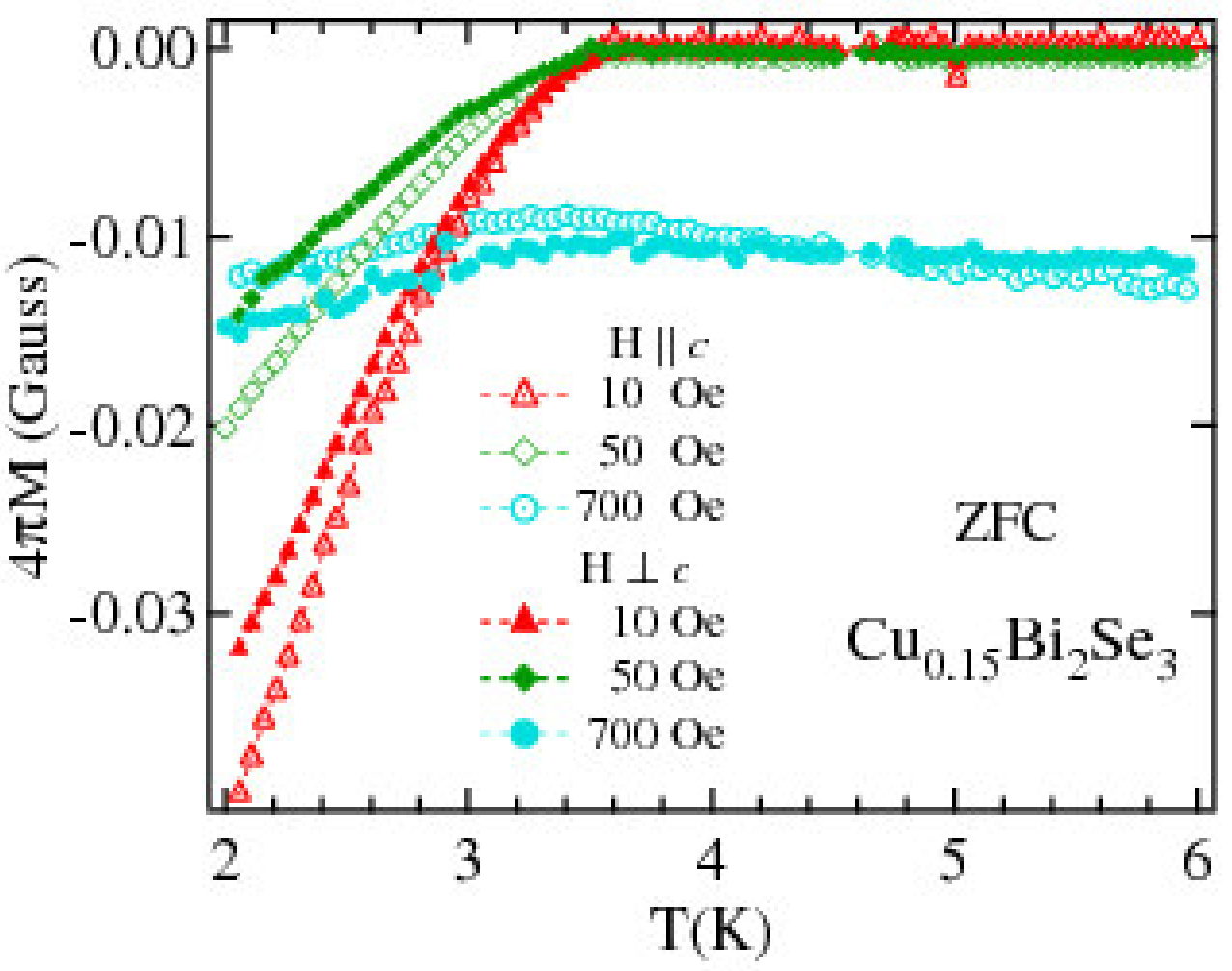

**Fig. 3**

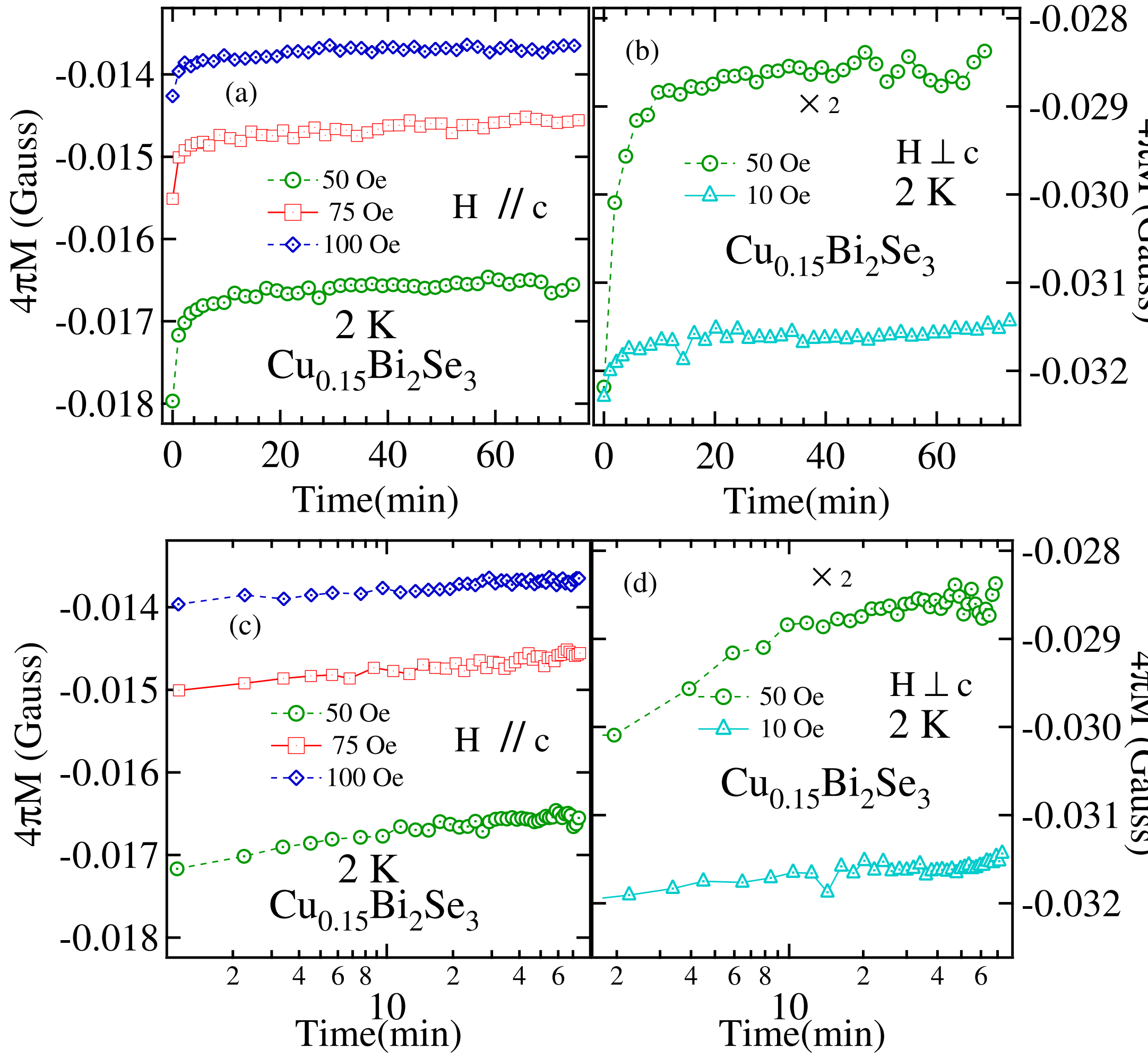

**Fig. 4**